\documentclass[fleqn,usenatbib]{mnras}

\usepackage{xcolor}
\usepackage{arydshln}
\usepackage{pdflscape}
\usepackage{lscape}
\usepackage{afterpage}
\usepackage{booktabs}
\usepackage{braket}
\usepackage{mathptmx,amsmath}
\usepackage{txfonts}

\usepackage[T1]{fontenc}
\usepackage{ae,aecompl}

\usepackage{graphicx}
\usepackage{tablefootnote,longtable,threeparttable}
\newcommand{\mearth}{M_\oplus}
\def\ms{\hbox{\,m\,s$^{-1}$}}         %m.s -1
\def\m2s2{\hbox{\,m$^{2}$\,s$^{-2}$}} %m2.s -2
\title{Expectations for the confirmation of Proxima\,c from a long-term radial velocity follow-up }

\author[M. Damasso and F. Del Sordo]{
M. Damasso,$^{1}$\thanks{E-mail: mario.damasso@inaf.it}
and F. Del Sordo$^{2,3}$\thanks{E-mail: fabiods@ia.forth.gr}\thanks{The authors equally contributed to this work}
\\
$^{1}$INAF - Osservatorio Astrofisico di Torino, Via Osservatorio 20, I-10025 Pino Torinese, Italy\\
$^{2}$Institute of Astrophysics, FORTH, GR-71110 Heraklion, Greece\\
$^{3}$Department of Physics, University of Crete, GR-70013 Heraklion, Greece
}

\date{Accepted XXX. Received YYY; in original form ZZZ}

\pubyear{2020}

\begin{document} 
\label{firstpage}
\pagerange{\pageref{firstpage}--\pageref{lastpage}}
\maketitle

\begin{abstract}
Proxima\,c, a candidate second planet orbiting Proxima Centauri, was detected with the radial velocity method. The announced long orbital period (5.21$^{+0.26}_{-0.22}$ years), and small semi-amplitude of the induced Doppler signal (1.2$\pm$0.4 $\ms$), make this detection challenging and the target worthy of a follow-up in the next years.
We intend to evaluate the impact of future data on the statistical significance of the detection through realistic simulated radial velocities to be added to the published dataset, spanning up to one orbital period of Proxima\,c in the time range 2019-2023. We find that the detection significance of Proxima\,c increases depending 
not only on the amount of data collected, but also on the number of instruments used, and especially on the timespan covered by the observational campaign.
However, on average we do not get strong statistical evidence and
we predict that, in the best-case scenario, in the next 5 years the detection of Proxima c can become significant at 4$\sigma$ level. If instead Proxima\,c does not exist, the detected signal may lower its significance down to 2$\sigma$.
\end{abstract}

\begin{keywords}
planetary systems -- methods: statistical -- star: individual: Proxima -- techniques: radial velocities
\end{keywords}

%
%-------------------------------------------------------------------
   
\section{Introduction}
   The discovery by \cite{anglada16} of the low-mass planet candidate Proxima\,b orbiting Proxima Centauri (Proxima hereafter) boosted the interest toward the nearest star to the Sun, and motivated the intensive follow-up campaign Red Dots\footnote{\url{https://reddots.space}} with the HARPS spectrograph to search for additional planetary companions. Observations from the Red Dots initiative were combined with archival data to cover a timespan of $\sim$17 years,
   %and the full time series of HARPS and UVES radial velocities (RVs) 
   and this led \cite[hereafter Paper I]{damasso19} to detect the candidate second planet Proxima\,c, with orbital period of 5.21$^{+0.26}_{-0.22}$ years and minimum mass 5.8$\pm$1.9 $\mearth$.
   This detection is challenging, and additional follow-up is necessary to confirm the planet. Among exoplanets with similar orbital period discovered so far using RVs, Proxima\,c holds the smallest Doppler signal with semi-amplitude 1.2$\pm$0.4 $\ms$, significant at a 3$\sigma$ level.
   
   Proxima\,c was discovered using a Gaussian process (GP) regression in a Bayesian framework, jointly modeling the planetary signals and the correlated term due to the stellar activity. In Paper I, the Bayesian evidence $\ln \mathcal{Z}$ of the models including one and two planets were compared to statistically support the detection of Proxima\,c, and it was found that the results depend on the choice of the prior ranges for the orbital parameters of the candidate planet. Therefore, one interesting issue to be explored is the influence of the priors on the results with an enlarged RV dataset. In this work we want to address this question by using realistically simulated RVs over a time span covering the period 2019-2023. The use of simulations to evaluate the expectations of a RV follow-up, featuring or not astrophysical contributions from stellar activity, is becoming common practice within large collaborations, where a long list of targets have to be monitored and an optimal observing strategy is crucial to cope with telescope time constraints. General scenarios were investigated, for instance, by \cite{2019DPSS}, while simulations were used for assessing the promises of RV follow-up for specific cases as GJ 1132 \citep[for current state-of-the-art spectrographs]{cloutier17}, K2-3 \citep[with HARPS-N]{damasso18}, Trappist-1 \citep[with SPIRou]{klein19}, K2-33 \citep[with SPIRou]{KD2020MNRAS}, Kepler-20 and K2-131 \citep[for current state-of-the-art spectrographs]{N2020AJ}. 
   We devised the simulations for the case of Proxima similarly to the cases studied by \cite{2019DPSS}. Through the simulations we aim at quantifying how much the two-planet model becomes statistically more favoured, and the precision and significance of the orbital parameters of Proxima\,c improve with new observations. Also, we verify whether in 5 years it will be possible to disprove the existence of Proxima\,c using only RVs.\\
   The results of this simulations can help in evaluating the effectiveness of long-term follow-up efforts with high-precision spectrographs for confirming such a long-period and small amplitude RV signal with higher significance, and to plan advantageous observing strategies. 
   So far, the detection via radial velocity of low-mass exoplanets on long orbital periods is a rather uncharted territory. In fact, Proxima \,c is the exoplanet with the longest period and a minimum mass in the super-Earth regime detected so far with the RV technique. Therefore, this work aims also at exploring the feasibility of detecting low-mass exoplanets with similar orbital periods, and the proposed  
   approach may be relevant for those that may be found in the future
   in this region of the parameter space.\\
   The paper is structured as follows. We begin by describing in Section \ref{sec:data} how we produced the simulated datasets. Then, in Section \ref{sec:mcanalysis} we present the method used for analysing the simulated datsets, and discuss the results of this analysis in Section \ref{sec:results}. We then draw some conclusions in Section \ref{sec:conclusions}.

%--------------------------------------------------------------------

\section{Building up the simulated datasets}
\label{sec:data}
\begin{figure}
   \centering
   \includegraphics[width=7.3cm]{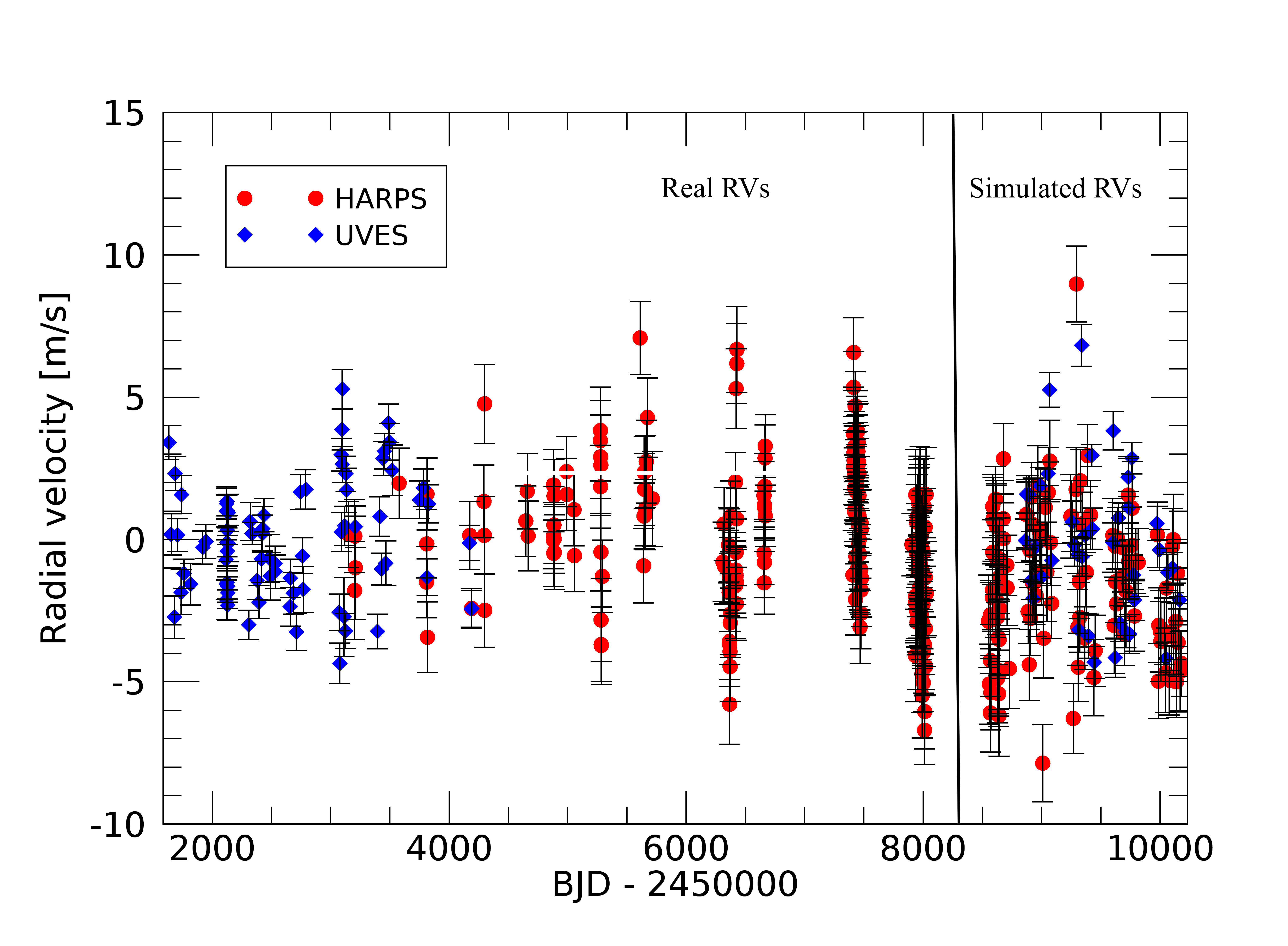}
   \includegraphics[width=7.3cm]{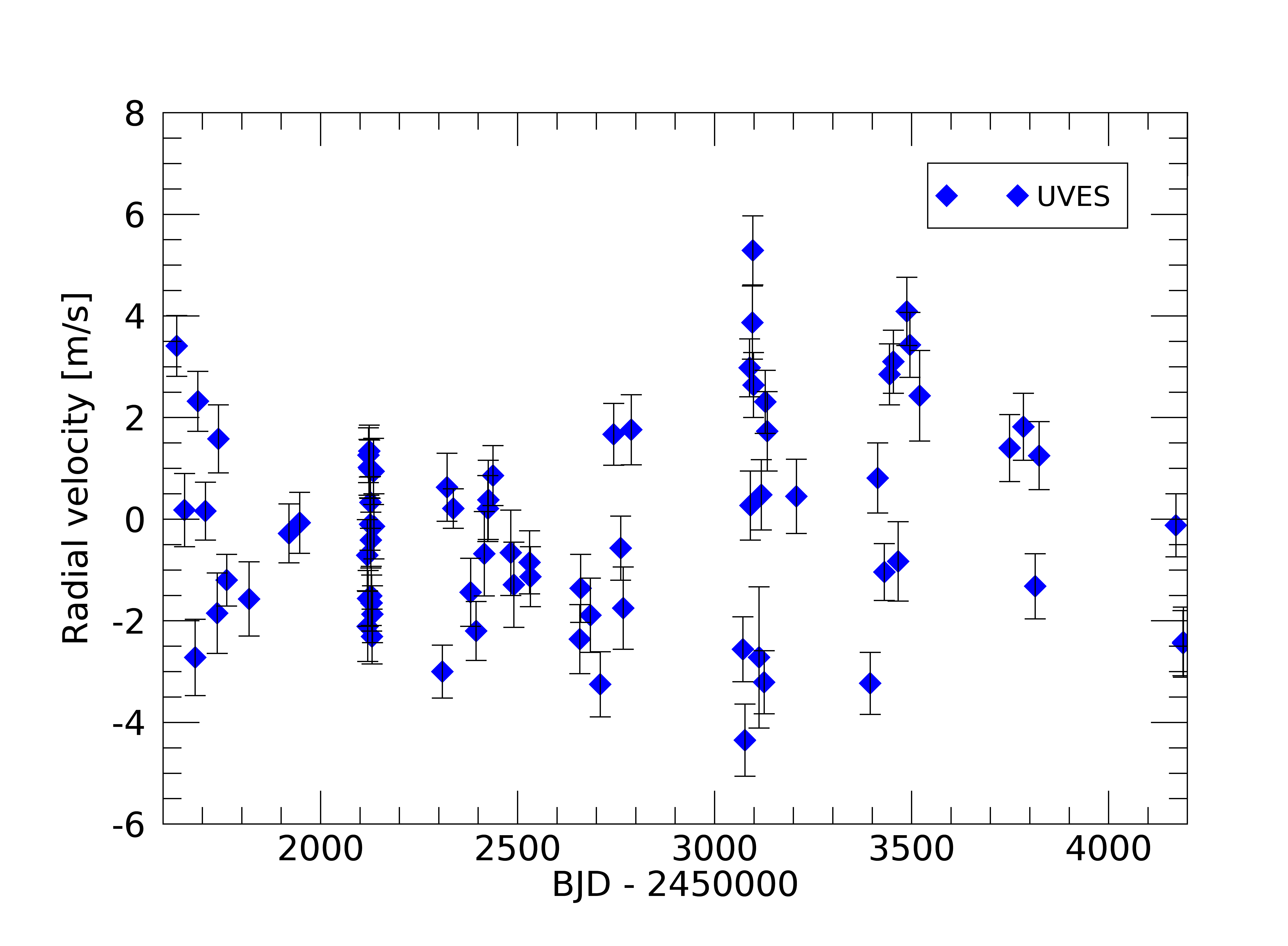}
    \includegraphics[width=7.3cm]{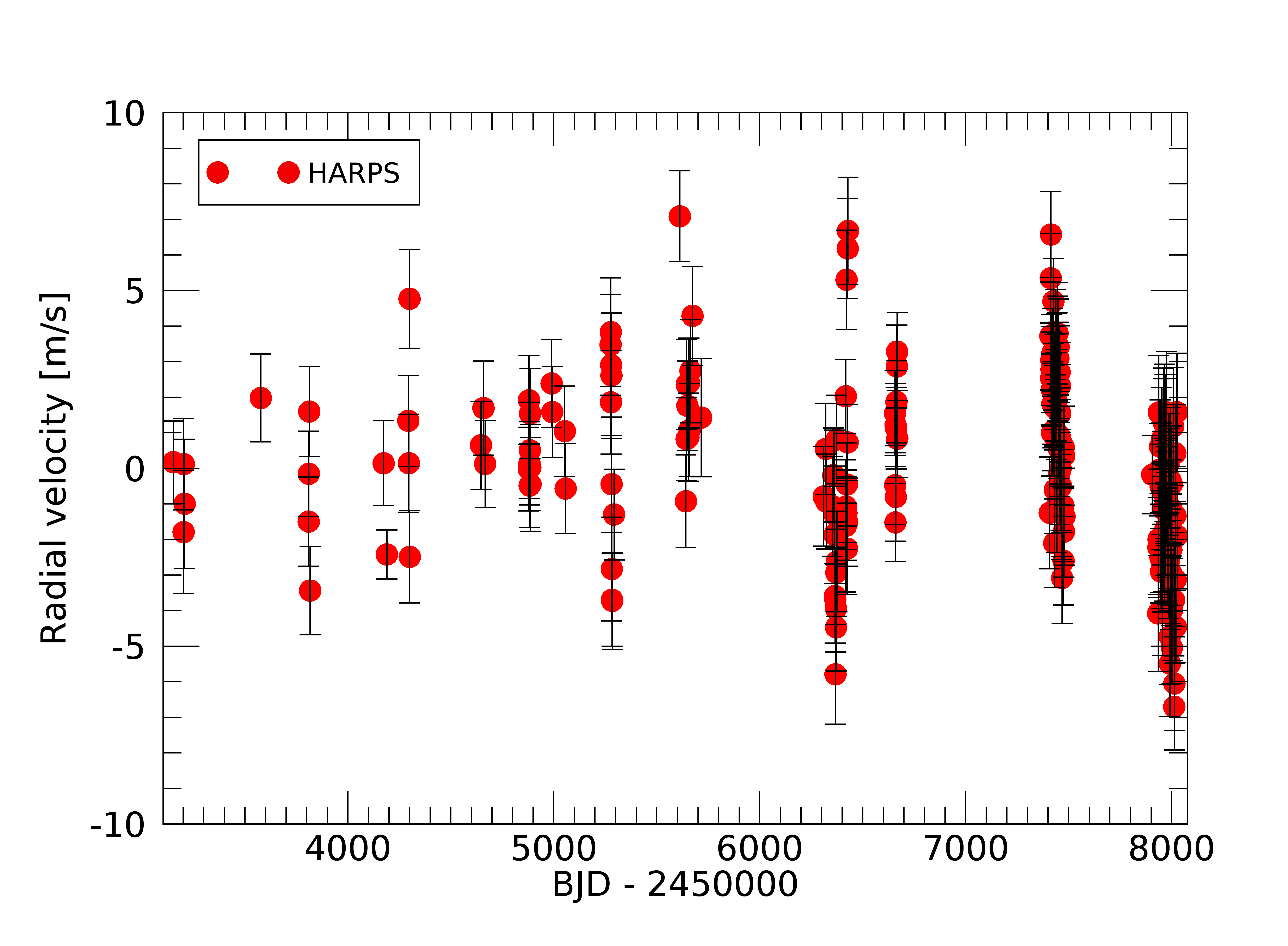}
    \includegraphics[width=7.3cm]{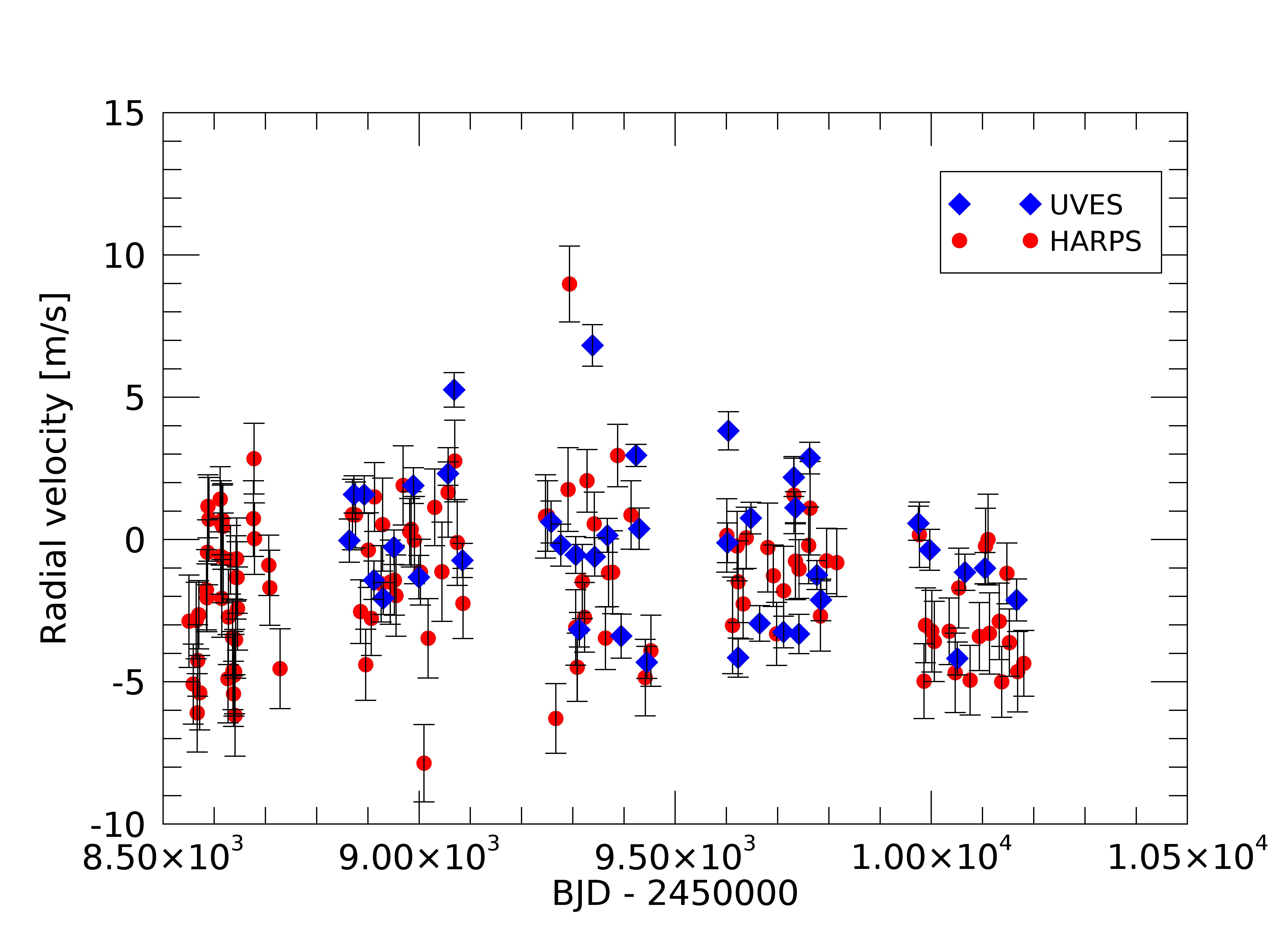}
      \caption{\textit{Upper panel:} Example of RV dataset for the Case \textit{H19}+All 20-23 (real and simulated data). \textit{Second panel:} Real UVES RVs used in Paper I.  \textit{Third panel:} Real HARPS RVs used in Paper I. \textit{Lower panel:} Zoomed-in view of the simulated data.
      The Generalized Lomb-Scargle periodogram of this specific dataset shows the highest peak at 1920 days.}
         \label{Fig:rv_simu_example}
   \end{figure}

We simulate RV datasets of Proxima obtained with HARPS and UVES, which are the instruments used to detect Proxima\,b and c, conceiving different scenarios. For each of them we simulate 20 datasets (Table \ref{Table:datasets}).
Although the proposed scenarios are arbitrary, they are based on past observational campaigns and on what we deem an affordable observing strategy to in-depth probe Proxima's planetary system, with particular regard to Proxima\,c. 
The longest time span we take into account is 2019-2023 so to include one additional orbit of Proxima\,c after the last published RVs. We also simulate time series over shorter periods, so to assess the potential of the RV technique before the full Gaia astrometric catalogue will be published, since Gaia astrometry is expected to confirm the existence of Proxima\,c and provide its real mass (Paper I).
Each simulated dataset is then added to the available 279 RVs obtained with UVES and HARPS and used in Paper I to discover Proxima\,c. These final datasets, consisting of real and simulated RVs, are then analysed, as described in Section \ref{sec:mcanalysis}.
\\
Proxima was observed in 2019 with HARPS (program 1102.C-0339(A); the RVs were not yet publicly available when this work was carried out), and we used the epochs of these observations, as found on the ESO data archival website, to generate mock RVs for the year 2019 (see \textit{H19} in Table \ref{Table:datasets}).
The archive reports 57 observations taken during 39 nights from 7$^{\rm th}$ March until 2$^{\rm nd}$ September 2019. For consistency with the real data analysed in Paper I, we binned the epochs on a nightly basis so to have one RV for each night, resulting in 39 time stamps. 
These epochs are included in all the simulated scenarios in Table \ref{Table:datasets}. \\
As for the epochs simulated after 2019, they are different for each dataset of a specific scenario, so that the results are not bound to a particular sampling. Each array of epochs is selected as a subset of the list of good epochs determined applying the following constraints: i) the angular distance of Proxima from the Moon greater than 35$^{\circ}$, and the Moon illumination less than 90$\%$; ii) the altitude of Proxima above the horizon greater than 35$^{\circ}$; iii) the Sun altitude less than -18$^{\circ}$ (astronomical dusk/dawn).\\    
We underline that, by construction, the simulated RVs must be considered as the average values of multiple observations per night, in analogy with the real dataset analysed in Paper I. This means that the actual observing time would likely be longer than one visit per night.\\
The uncertainties $\sigma_{\rm RV}(t)$ of the simulated RVs at epoch $t$ are randomly generated using the real measurements in Paper I,  adopting a normal distribution with mean equal to the median of the $\sigma_{\rm RV}(t)$ for each instrument (0.6 and 1.3 $\ms$, for UVES and HARPS respectively), and their RMS as the dispersion of the probability distribution. 
As a further constraint, the values of $\sigma_{\rm RV}(t)$ are all kept in the range 0.35-0.8 \ms and 1-2 \ms, for UVES and HARPS simulated data respectively, which are the intervals where the $\sigma_{\rm RV}(t)$ of the real data generally fall. We kept the UVES sampling more sparse than that of HARPS taking into account the higher precision of the UVES data, and that accessing to UVES for the purposes of a dense follow-up has historically proven to be more difficult. We show in Fig. \ref{Fig:rv_simu_example} an example of simulated RV dataset that includes both HARPS and UVES data up to the year 2023.\\
We want to answer two questions: how will the results from a longer term follow-up look like in the hypothesis that Proxima\,c exists? What if it does not exist? To answer, we conceived two sets of simulations. For the first (hereafter \textit{problem 1}), we assume that the solution proposed in Paper I is correct. This means assuming that the orbits of Proxima\,b, Proxima\,c and the stellar activity pattern are well described by the parameters listed in Table \ref{tab:sim2planets} (column 2). We use them and their uncertainties to randomly generate an array of parameters for each dataset (semi-amplitudes $K_{\rm b}$ and $K_{\rm c}$, orbital periods $P_{\rm b}$ and $P_{\rm c}$, and times of inferior conjunction $T_{\rm 0,\:b}$ and $T_{\rm 0,\:c}$ of the planetary orbits; hyper-parameters $h$, $\lambda$, $\theta$, and $w$ for the correlated signal due to the stellar activity), from which we calculate the \textit{exact} RV$(t)$ at each selected epoch $t$. 
As done in \cite{2019DPSS} to generate fully mock datasets, the term due to the activity of Proxima, $\Delta RV_{\rm act}(t)$, was randomly drawn using the \texttt{sample} function of the \texttt{GP} object implemented in the \texttt{GEORGEv0.2.1} package \citep{george15}, which was used in Paper I to define the GP framework and fit the data (see also Section \ref{sec:mcanalysis}). Once we fix a set of GP hyper-parameters randomly drawn from the posterior distributions determined in Paper I, the \texttt{sample} function returns a randomly drawn list of predictions of $\Delta RV_{\rm act}(t)$ for each simulated epoch. The definitive RV($t$) are obtained by randomly shifting the \textit{exact} values within the error bar given by the sum in quadrature of $\sigma_{\rm RV}(t)$ and an uncorrelated jitter term $\sigma_{\rm instr.,\: jit}$ (one for each instrument, adopting the best-fit results of Paper I), using a normal distribution centred on zero and with $\sigma=\sqrt{ \sigma_{\rm RV}(t)^2+\sigma_{\rm instr.,\: jit}^2}$. Finally, we added a constant offset to the mock RVs (one for each instrument), using the best-fit values of Paper I.

With a second set of simulations (hereafter \textit{problem 2}), we want to  to verify whether new RVs spanning 5 years may be sufficient to disprove the existence of Proxima\,c.
To this purpose, we simulate datasets as for \textit{problem 1}, but including only the signals due to Proxima\,b and stellar activity, using the best-fit solution for the corresponding one-planet model derived in Paper I (Table \ref{tab:sim1planet}). 

A note of caution needs to be added. We worked under the hypothesis that the stellar noise as modeled in Paper I keeps its structure and properties for the whole simulated campaign. We used the results of Paper I to predict the values of $\Delta RV_{\rm act}(t)$, as described above. As the RV signal originated by stellar activity is by its very nature stochastic, of course it is difficult to be predicted over a long time range, especially for fully convective stars such Proxima, with quite intense activity. Nonetheless, our activity model was obtained by analysing more than 17 years of RVs, and in Paper I it was demonstrated how it would not change much even if the latest HARPS observational campaigns were not included in the analysis. Therefore we deem reasonable to assume our activity model valid over 5 more years.

\subsection{Quantifying the observational effort}
On the basis of previous observational campaigns, and by using information in the ESO archive, we can quantify  the  true  investment in observational time and how this relates to the investment already made.
First of all, we know that the dataset \textit{H19} required 14.3 hours of telescope time. Each spectrum was obtained with an exposure of 900 seconds, whilst only one of the 57 observations required an exposure of 1200 s.
Moreover, as said, after binning over single nights, 57 spectra resulted in 39 RVs, hence providing a conversion factor of about 1.5.  
Previous HARPS spectra had exposure times varying from 450 s up to 1200 s. In particular, spectra obtained within the Red Dots campaign were characterized by a 1200 s exposure time. Therefore we can take 1200 s as a conservative estimate for the average exposure time required for a single spectrum, excluding overheads. We also consider that an average of 1.5 spectra are needed to produce an RV. We deem this a rather conservative estimate of the telescope time needed to produce datasets similar to those we here suggest.
As for observations carried out with UVES, in agreement with previous campaigns, we consider that a single RV is extracted by three spectra taken in a single night, each of them with an exposure of 800 s, excluding overheads. This gives a total of 2400 s per RV.
In the last column of Table \ref{Table:datasets} we provide an estimate of the required observational effort for each scenario.
For comparison, the Red Dots campaign required 66 spectra to produce 61 RVs, for a total telescope time of 22 hours in a single season.

\begin{table*}
  \caption[]{Description of the simulate scenarios. Each of them is added to the real 279 RVs analysed in Paper I. The array of simulated epochs is different for every dataset within each scenario.
  For each scenario we provide a description, the  number of simulated data $N_{\rm RV,\: sim}$, the total number of data analysed $N_{\rm RV,\: tot}$ (279 real RVs + simulated RVs), the total timespan in years, and an estimate of the actual needed observational time $T_{obs}$ in hours for data in the time range 2020-2023 (overheads not included). For the scenario \textit{H19}, related to observation that have been already carried out, we already know from ESO archive that $T_{obs}=14.3$. For all the other datasets, $T_{obs}$ is therefore the sum of 14.3 and the additional required telescope time based on our defined sampling. }
         \label{Table:datasets}
         \centering
         %\scriptsize
   \begin{tabular}{lp{0.5\textwidth}cccc}
            \hline
            \hline
            \noalign{\smallskip}
            \textbf{Scenario}  & \textbf{Description}& $N_{\rm RV,\:sim}$ & $N_{\rm RV,\:tot}$ & Timespan & $T_{obs}$ \\
            \noalign{\smallskip}
            & & & & [years] & [hours] \\
            \noalign{\smallskip}
            \hline
            \noalign{\smallskip}
            \textit{H19} & Includes real epochs of observations of Proxima carried out with HARPS (program 102.C-0339(A)) from March 7$^{th}$ through September 2$^{nd}$ 2019, for a total of 57 spectra spanning 39 nights. We considered N=39 epochs to simulate nightly binned RVs. Since this is a set of epochs corresponding to observations actually performed, it is included in all the following datasets. & 39 & 318 & 19.5 & 14.3\\
            \noalign{\smallskip}
            \textit{H19}-23 & HARPS simulated data for epochs during 2020-2023.
            For each of the 4 years and for every simulated RV dataset, first we randomly selected one epoch per week satisfying our assumed observational constraints. Then, we randomly selected 80 epochs out of the 32*4=128 total sample. An average number of 20 RV/year can be considered as a conservative estimate for a real follow-up campaign with HARPS focused on Proxima\,c. & 119 & 398 & 23.5 &14.3 +40\\
            \noalign{\smallskip}
            \textit{H19}+U20-23 & 
         Simulated UVES measurements during 2020-2023. For each of the 4 years and for every simulated RV dataset, first we randomly selected one epoch per week among those satisfying our assumed observational constraints. Then, we randomly selected 40 epochs out of the 32*4=128 total sample. An average number of 10 RV/year can be considered as a conservative estimate for a real follow-up campaign with UVES focused on Proxima\,c. & 79 & 358 & 23.5 & 14.3 + 26.7 \\
            \noalign{\smallskip}
            \textit{H19}+All 20-23 & A combination of datasets  \textit{H19}-23 and \textit{H19}+U20-23. & 159 & 438 & 23.5 & 14.3 + 66.7\\
            \noalign{\smallskip}
           \textit{H19}-23 RV$^{+}$ & Same as \textit{H19}-23, but instead of having
           20 RV/year on average, here we increase the average number to 40 RV/year. This sample is randomly selected within the total of good epochs per year according to the observational constraints. 
           The total of 160 simulated RVs in the 4-year time span is 
           less conservative than the previous scenario, nonetheless still realistic. & 199 & 478 & 23.5 & 14.3 + 80\\
            \noalign{\smallskip}
             \textit{H19}-21 & HARPS simulated data. 60 randomly selected epochs both in 2020 and 2021
            among those satisfying our assumed observational constraints. & 159 & 438 & 21.5 &14.3 + 60\\
            \noalign{\smallskip}
            \hline
            \hline
     \end{tabular}    
\end{table*}

\section{Analysis of the simulated datasets}
\label{sec:mcanalysis}
We analysed each dataset with models where the correlated noise is fitted with a quasi-periodic function described by
\begin{eqnarray} \label{eq:eqgpkernel}
k(t, t^{\prime}) = h^2\cdot\exp\bigg[-\frac{(t-t^{\prime})^2}{2\lambda^2} - \frac{sin^{2}(\dfrac{\pi(t-t^{\prime})}{\theta})}{2w^2}\bigg] + \nonumber \\
+\, (\sigma^{2}_{\rm instr,\: RV}(t)\,+\,\sigma_{\rm inst,\: jit}^{2})\cdot\delta_{\rm t, t^{\prime}}
\end{eqnarray}
This covariance matrix includes four hyper-parameters: the rotation period of the star, $\theta$; the decay timescale of the correlations, $\lambda$, related to the average lifetime of active regions; the length scale of the periodic component, $w$, describing the level of high-frequency variation within a complete stellar rotation; and, finally, the amplitude of the correlations, $h$.
The term $\sigma_{\rm instr.,\: RV}(t)$ is the error on the RV at time \textit{t} for each instrument (namely, UVES, and HARPS, treated as a different instrument for pre-2016 and for post-2016 observations); $\sigma_{\rm instr.,\: jit}$ is a constant uncorrelated jitter term (one for each instrument), introduced to take into account other sources of noise not included in $\sigma_{\rm RV}(t)$, and $\delta_{t,t^{\prime}}$ is the Kronecker delta.
For \textit{problem I}, each time series was analysed using three different models also used in Paper I. First, we fit only one planet, Proxima\,b and calculate the Bayesian evidence: this can tell us how suitable a GP+one-planet regression is to describe the signal actually generated by Proxima \,c.
%how well
%the signal generated by Proxima \,c.
%the proposed model for stellar noise can explain the signal generated by Proxima \,c.} 
Second, we use a two-planet model, the same it was adopted in Paper I to estimate the orbital parameters of Proxima \,c. Third, we employ a two-planet model with a large prior on the semi-amplitude, period, and epoch of inferior conjunction of the Doppler signal of Proxima \,c. This last analysis aims to quantify the dependence of the Bayesian evidence $\ln\mathcal{Z}$ on the choice of the priors. We use the same free parameters and priors as in Paper I (see Table \ref{tab:sim2planets}), therefore taking no advantage of any a priori information from Paper I.

For \textit{problem II} we performed the same analysis as for \textit{problem 1}, but we do not calculate the Bayesian evidence of a model with a large prior as we are interested only in seeing whether a long-period signal similar to that reported in Paper I is still present.
Moreover, we calculated Generalized-Lomb-Scargle periodograms \citep[GLS;][]{zech09} of the simulated datasets as additional diagnostics. We do so because in Paper I the signal of Proxima\,c clearly emerged as the main peak in the GLS periodogram of the RV residuals after subtracting from the original data set the
best-fit spectroscopic signal induced by Proxima\,b. Here, we are interested in investigating how the periodograms change under the hypothesis that planet c is not there. 

We use the quantities described in Table \ref{Table:analysedparam} to quantify our findings. They all are either ratios between injected values and retrieved ones, or ratios between a retrieved quantity and its lower uncertainty, so to quantify its significance in terms of $\sigma$-level. Also, we quantify the difference of the Bayesian evidence $\ln\mathcal{Z}$ between the different models used to analyse the mock datasets.

All the analyses were carried out with the open source Bayesian inference tool \texttt{MultiNestv3.10} (e.g. \citealt{feroz13}), through the \texttt{pyMultiNest} \texttt{python} wrapper \citep{buchner14}, with 500 live points and a sampling efficiency of 0.5, as done in Paper I. The GP module is the same publicly available \texttt{GEORGEv0.2.1} \texttt{python} library used to generate the datasets.  

%--------------------------------------------------------------------
\section{Results and discussion}
\label{sec:results}
\subsection{\textit{Problem 1}}
The results for \textit{problem 1} are presented in detail in Table \ref{tab:sim2planets}.
First, we notice that the observations carried out with HARPS in 2019 are not expected to bring a significant improvement in the Bayesian evidence difference $\Delta \ln\mathcal{Z}_(2pl-1pl)$, and the significance of the planetary signal can be only slightly larger than 3$\sigma$, which is the result already obtained in Paper I. Still, this will not be sufficient to confirm the existence of Proxima c. Similar results are obtained in the case of the H19-21 scenario, which includes 120 more RVs than H19.
We find that the hyper-parameters are in general well recovered. $K_{\rm c}$ is generally well recovered too, apart for the case H19+All 20-23, where the retrieved value is on average underestimated at 2$\sigma$ level. This result seems to indicate that the combination of data coming from two instruments may occasionally be disadvantageous, although producing a more dense time series. The best case scenario is that of high-cadence sampling with only one instrument, H19-23 RV$^{+}$, where the signal becomes significant at 4$\sigma$ and the Bayesian evidence indicates the two-planet model is almost 25 times more probable than the one-planet model if the analysis is carried out with a narrow prior. This must have to do with a better fit of the correlated quasi-periodic stellar activity term. Nonetheless, there is no significant improvement with respect to the H19-23 case, and it comes at the cost of a doubled observational effort (40 versus 20 RVs/year). 
We also notice that spreading observations in 5 years (H19-23, 119 additional data points) is a better strategy than concentrating many observations in three years (H19-21, 159 additional data points). This points out that following one entire orbit of Proxima\,c brings more information on its orbital parameters with respect to a dense monitoring limited in time, as shown by the increased significance $P_{\rm c,\:50\%}$/$\sigma^{\rm -}_{\rm P_{\rm c}}$ of the retrieved orbital period.
Concerning our statistical analysis, two main results are that i) on average, we should not expect to get strong statistical evidence of the two-planet model over the one-planet model ($\langle\ln \mathcal{Z}(2p_w)$ - $\ln \mathcal{Z}(2p_n)\rangle$<5 for all the scenarios, adopting the scale for the interpretation of model probabilities given by \citealt{feroz11}); and ii) for all the scenarios under investigation the results are less influenced by the choice of the priors of the parameters of planet c. In fact, the difference $\ln \mathcal{Z}(2p_w)$ - $\ln \mathcal{Z}(2p_n)$ between the Bayesian evidences of the two-planet models, with large and narrow priors, is on average more than halved with respect to Paper I.   

\subsection{\textit{Problem 2}}
\begin{figure}
   \centering
     \includegraphics[width=\hsize]{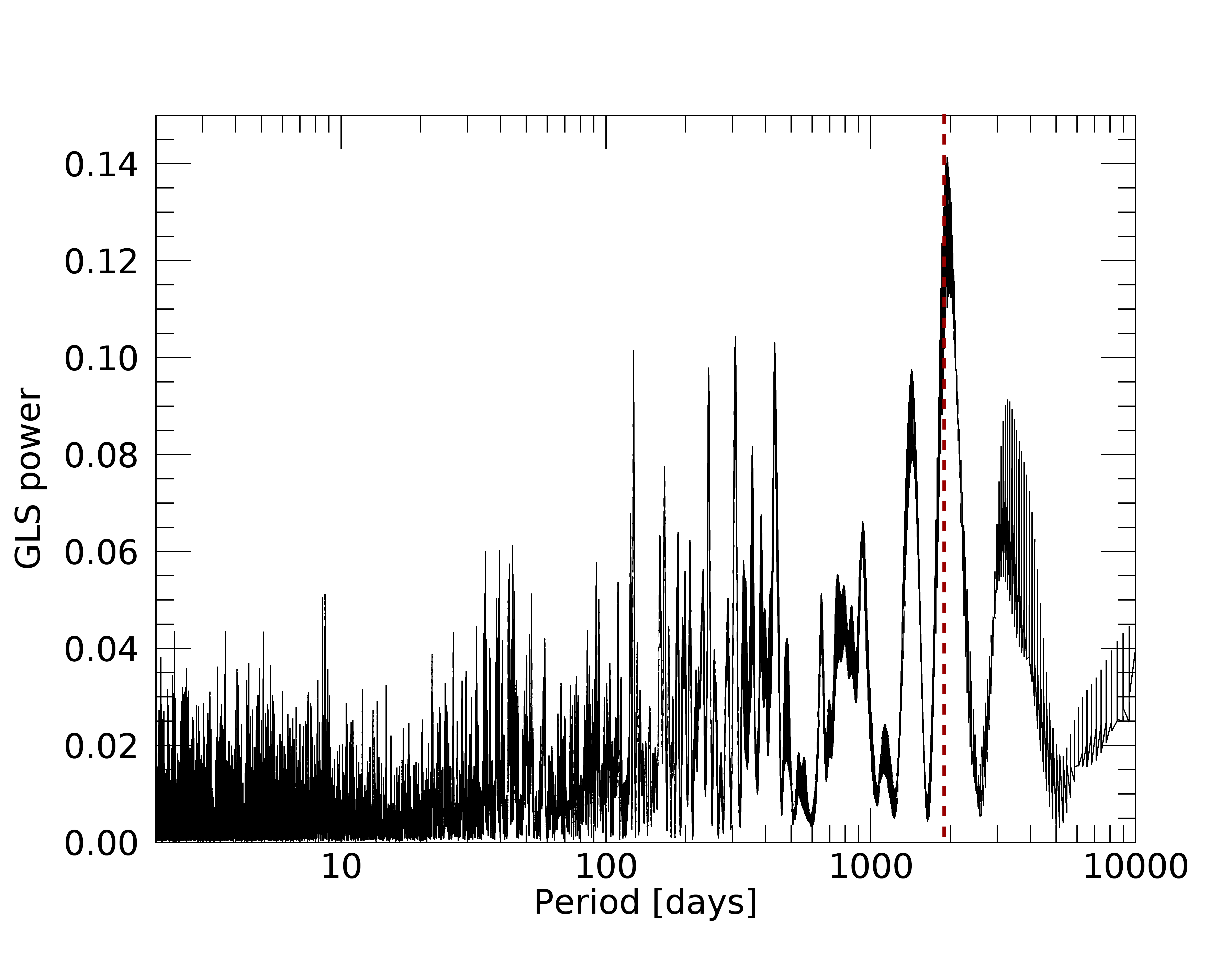}
   \includegraphics[width=\hsize]{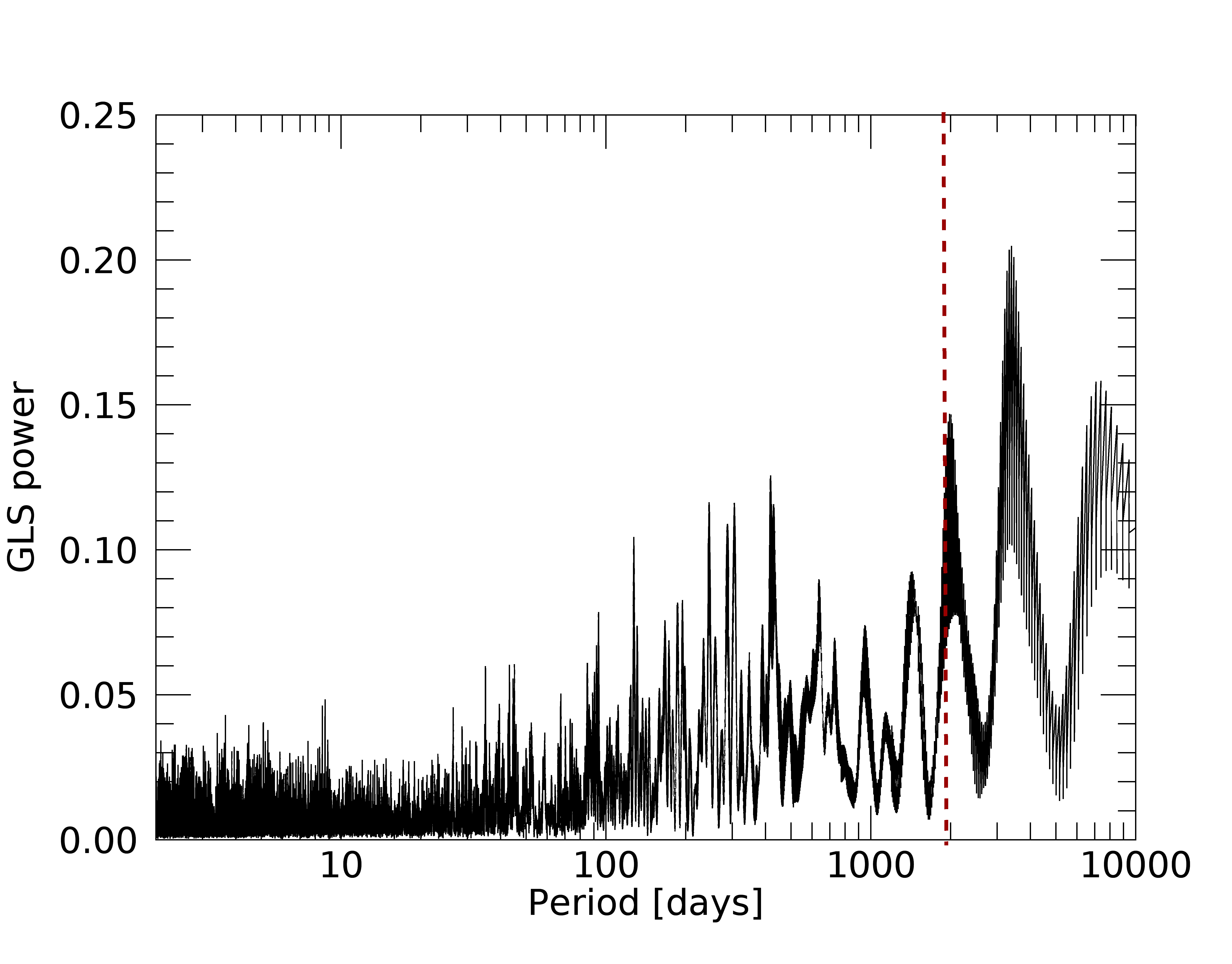}
    \includegraphics[width=\hsize]{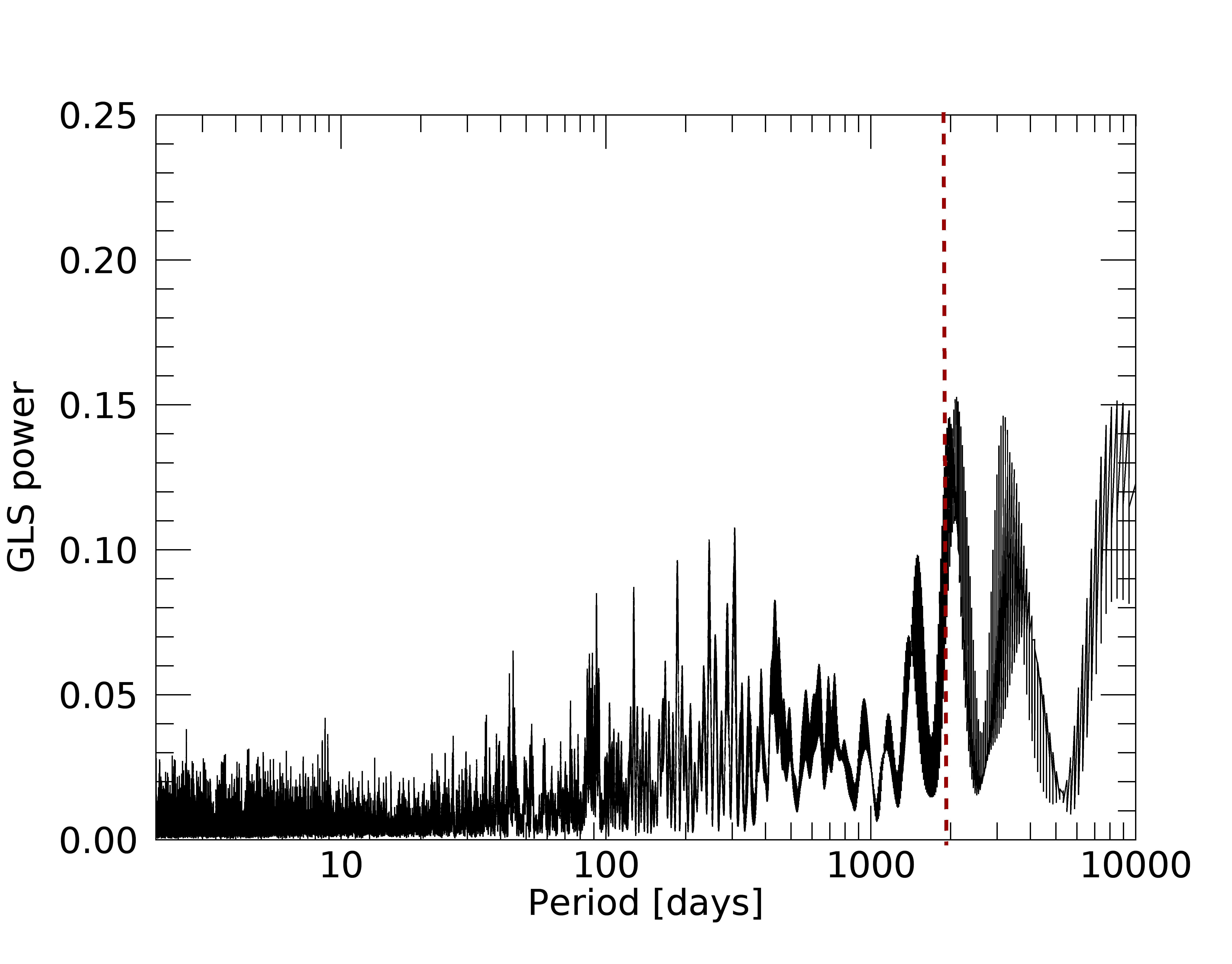}
      \caption{Averaged GLS periodograms of the residuals for data of \textit{problem 2}, after removing from each simulated dataset only the best-fit solution for the planet Proxima\,b, as derived from fitting a GP+1 circular planet model. The dashed red line marks the period of the candidate Proxima\,c derived in Paper I. \textit{Upper panel:} Case \textit{H19-23}. \textit{Middle panel:} Case \textit{H19+U20-23}. \textit{Lower panel:} Case \textit{H19+All 20-23}.}
         \label{Fig:period_mean}
   \end{figure}

Table \ref{tab:sim1planet} summarizes the results about \textit{problem 2}. 
We opt for analysing only three of the datasets in Table \ref{Table:datasets}. We exclude those not covering 5 years (\textit{H19} and \textit{H19-23})and that with high cadence (\textit{H19-23 RV$^{+}$}).
The reason for this choice is that, by starting our analysis from the scenario \textit{H19+All 20-23}, we saw that the one planet model is statistically as significant as the two planet model.
Moreover, we learn from \textit{Problem 1} that the timespan covered by the observations affects the results more than the total number of RVs.  As a consequence, we argue that analyzing scenarios that include shorter timespan (\textit{H19} and \textit{H19-21}) or very high cadence (\textit{H19-23 RV$^{+}$}) would not change much the results coming from the analysis of \textit{H19+All 20-23}. Nonetheless, for completeness we decided to carry out also the analysis of \textit{H19-23} and \textit{H19+U20-23} to study the possible dependence on the used instrument.
In the three analysed scenarios, a signal with period of $\sim1900$ days still appears, but its significance substantially lowers on average (1.9$\sigma$ in the lowest case). The Bayesian evidences for the one-planet and two-planet models become equivalent. 
We then calculate the GLS periodograms of the RV residuals, obtained after removing the best-fit solution for Proxima\,b's induced signal.
From these we derive the average periodograms for each scenario using bins of five consecutive frequencies (due to their different sampling, the periodograms of each scenario do not necessarily posses the same frequency stamps). They are shown in Fig. \ref{Fig:period_mean}. Even though, by construction, these periodograms are not statistically rigorous, they carry general information about the average properties of single periodograms and they are indicative of whether or not a peak appears in many periodograms. For the three scenarios we can still see a peak with a period similar to that of Proxima\,c, although for the H19-23 and H19+All20-23 cases this peak has a lower GLS power with respect to Paper I, whilst for H19+U20-23 it is not anymore the most significant peak, which now falls at the alias frequency due to the time span of the observations. This looks consistent with analysis presented in Table \ref{tab:sim1planet}. 

\begin{figure}
   \centering
   \includegraphics[width=\hsize]{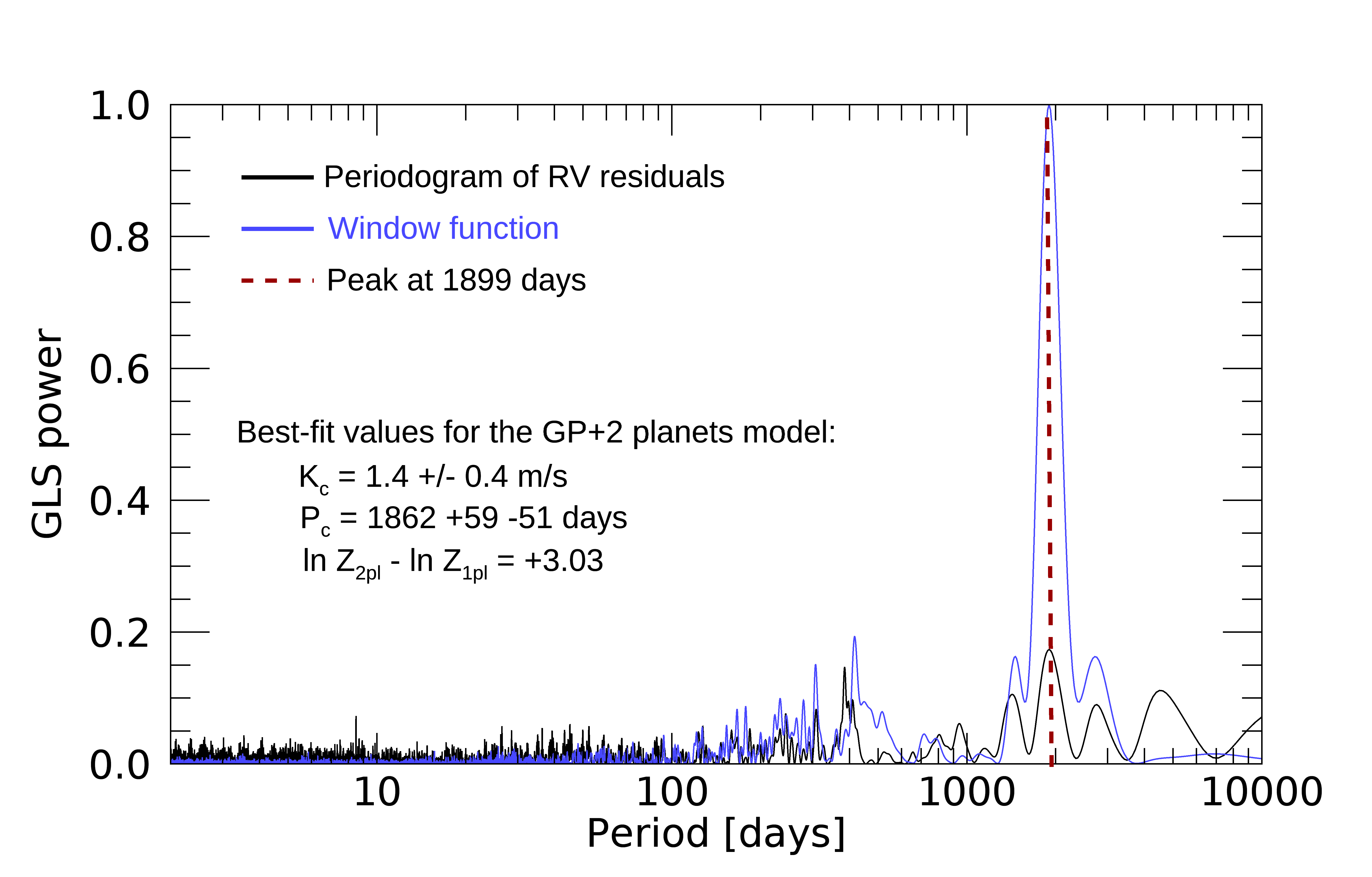}\\
   \includegraphics[width=\hsize]{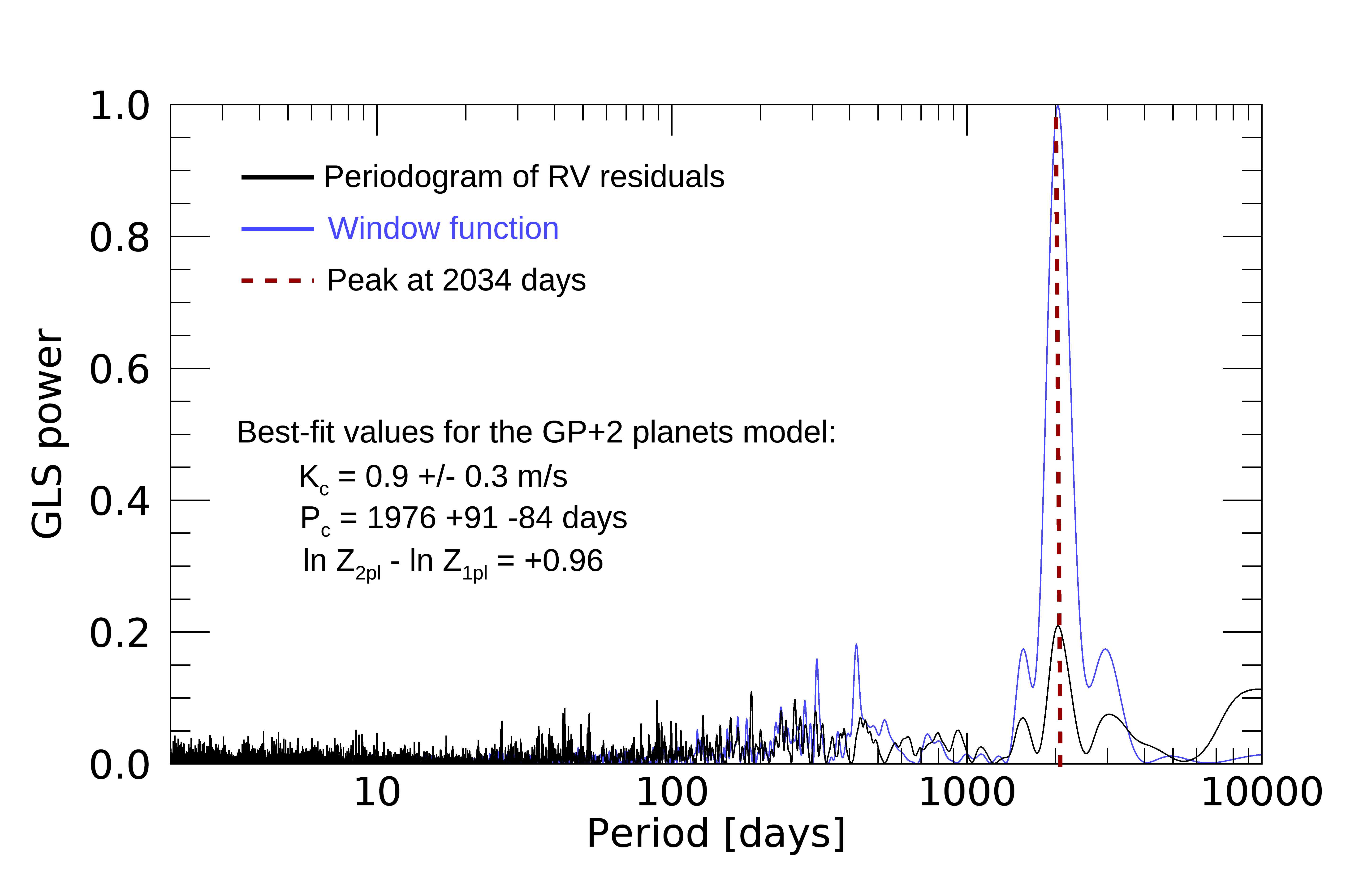}\\
   \includegraphics[width=\hsize]{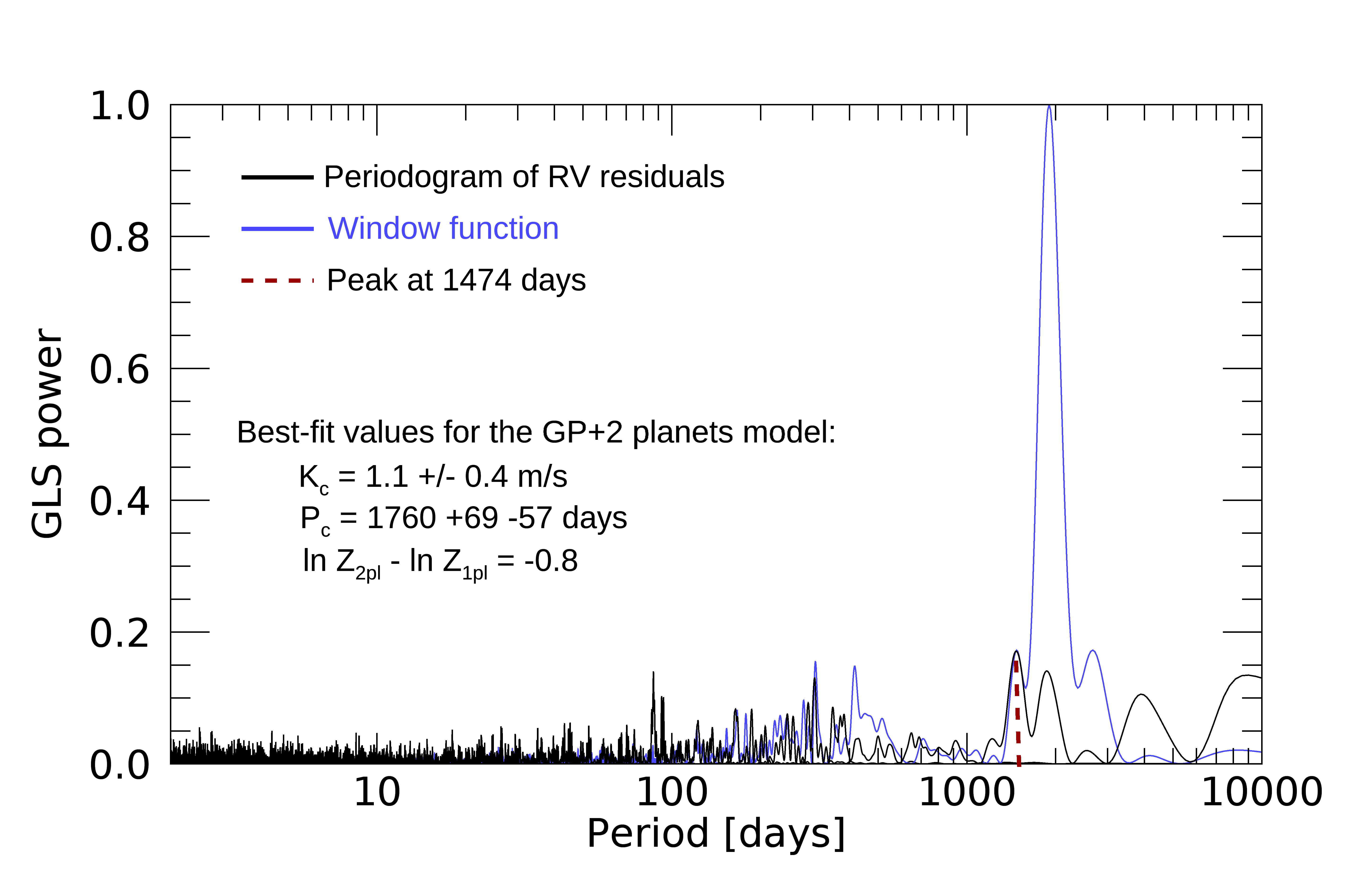}\\
   \caption{GLS periodograms (black lines) of RV residual time series for selected mock datasets that do not contain the Doppler signal induced by Proxima\,c, with the corresponding window function overplotted (blue line). The peak at lowest frequency in the window function is shifted in frequency to show the location of the frequency aliases related to the data time span that correspond to periods close to the orbital period of Proxima\,c. \textit{Upper panels}. Scenario \textit{H19-23}. \textit{Middle and lower panels}. Scenario \textit{H19+All 20-23}.
              }
         \label{Fig:gls_single_period}
\end{figure}

For illustrative purposes, we show in Fig. \ref{Fig:gls_single_period} some examples of GLS periodograms related to \textit{problem 2} and calculated for RV residuals derived from simulated measurements including only the signals of Proxima\,b and stellar activity. These periodograms are representative of those that could be actually obtained after a long-term follow-up of Proxima. The residuals are those calculated by subtracting only the best-fit solution for the spectroscopic orbit of Proxima\,b, obtained by fitting a model which includes the orbit of Proxima\,b and a GP quasi-periodic kernel to model the stellar activity correlated signal.

The periodogram in the upper panel has a clear peak exactly at the orbital period of Proxima\,c, and resembles to the corresponding periodogram obtained for real data (see Fig. 1 in Paper I). Even though the spectroscopic orbit of Proxima\,c was not injected in the simulated HARPS RVs, the signal of the candidate planet dominates the residuals. This is an interesting case, since $\ln \mathcal{Z}_{2 \:planets}$ - $\ln \mathcal{Z}_{1 \:planet}$=+3, making the detection of planet c even more robust than in Paper I. Consequently, the MC fit (GP + two planets model) retrieves a signal of semi-amplitude $K_{\rm c}$ more than 3$\sigma$ significant, and with an orbital period more accurate than that determined in Paper I (Table \ref{tab:sim2planets}). 

The periodograms in the other two panels refer to dataset from the H19+All 20-23 scenario. The first illustrates a case where the signal of Proxima\,c is still that with the highest power, but the retrieved semi-amplitude $K_{\rm c}$ (GP + two planets model) is less than 1 $\ms$. Last panel of Fig. \ref{Fig:gls_single_period} illustrates a case where the peak of the periodogram does not occur close to the orbital period of Proxima\,c. For this specific dataset, the detection of planet c is less significant, since the model including only the orbit of Proxima\,b is more robust ($\ln \mathcal{Z}_{2 \:planets}$ - $\ln \mathcal{Z}_{1 \:planet}$ < 0), and the best-fit orbital period $P_{\rm c}$ shifts toward the lower edge of the prior, since the peak of the periodogram occurs at 1474 days.  

\begin{table}
  \caption[]{Figures of merit to assess the expectations for each scenario described in Table \ref{Table:datasets}. These quantities are evaluated for each mock RV dataset taking into account the best-fit solution for the GP+2 planets regression analysis.}
         \label{Table:analysedparam}
         \centering
         \scriptsize
   \begin{tabular}{lp{0.35\textwidth}}
            \hline
            \hline
            \noalign{\smallskip}
            \textbf{Quantity}  & \textbf{Description} \\
            \noalign{\smallskip}
            \hline
            \noalign{\smallskip}
            $h_{\rm ratio}$ & Ratio between the retrieved amplitude of the activity term and the corresponding injected value;\\
            \noalign{\smallskip}
            $\lambda_{\rm AR, ratio}$ & Ratio between the retrieved active region evolutionary time scale and the corresponding injected value;\\
            \noalign{\smallskip}
            $w_{\rm ratio}$ & Ratio between the retrieved value of the $w$ hyper-parameter and the corresponding injected value;\\
            \noalign{\smallskip}
            $P_{\rm rot,\: ratio}$ & Ratio between the retrieved stellar rotation period in the activity term and the corresponding injected value;\\
            \noalign{\smallskip}
         %   $\sigma_{\rm ratio}$ & Average of the ratio between the total error budget in the retrieved model and the injected uncertainty (at epoch $t$) for each mock RV dataset (see Eq. \ref{eq_sigma_rat});\\
         %   \noalign{\smallskip}
            $K_{\rm c,\:ratio\: 50\%}$ & Ratio between the 50$^{\rm th}$ percentile of the retrieved planetary semi-amplitude  (the best-fit value of $K_{\rm b}$) and the corresponding injected semi-amplitude $K_{\rm c,\:inj}$=1.2 \ms;\\
         %   \noalign{\smallskip}
         %   $K_{\rm b,\:ratio\: 68.3\%}$ & Ratio between the 68.3$^{\rm th}$ percentile of the planetary semi-amplitude $K_{\rm b}$ retrieved from each mock dataset and the corresponding injected value;\\
            \noalign{\smallskip}
             $K_{\rm c,\:50\%}$/$\sigma^{\rm -}_{\rm K_{\rm c}}$ & Ratio between the 50$^{\rm th}$ percentile of the posterior of the planetary semi-amplitude $K_{\rm c}$ and its lower uncertainty. We assume this parameter to quantify the significance of the retrieved planetary Doppler semi-amplitude with respect to a null value;\\
            \noalign{\smallskip}
             $P_{\rm c,\:ratio\: 50\%}$ & Ratio between the 50$^{\rm th}$ percentile of the  retrieved planetary orbital period and the corresponding injected orbital period $P_{\rm c},P_{inj}$=1900 days;\\
            \noalign{\smallskip}     
            $P_{\rm c,\:50\%}$/$\sigma^{\rm -}_{\rm P_{\rm c}}$ & Ratio between the 50$^{\rm th}$ percentile of the posterior of the planetary orbital period $P_{\rm c}$ and its lower uncertainty. We assume this parameter to quantify the significance of the retrieved planetary orbital period.\\
             \noalign{\smallskip}
            \hline
            \hline
     \end{tabular}    
\end{table}

%\clearpage

%--------------------------------------------------------------------

%\begin{scriptsize} 
\begin{landscape}
\begin{table}
%\begin{table}
       \caption{Results for \textit{Problem 1}. We use the GP regression analysis applied to RVs, including two circular orbital equations, for the different simulated datasets. The second column reports results of Paper I for comparison. These were used to generate the mock RVs. The other columns report results for the figures of merit of Tab. \ref{Table:analysedparam} averaged over all the analysed mock data. In the last 3 rows we report the difference in Bayesian evidences coming from three different proposed models. These are the difference between two-planet and one-planet model, with a large prior adopted for Proxima\,c; then, the same, but with a narrow prior on Proxima\,c which is the same adopted in Paper I for estimating the best-fit system parameters; and in the last row we report the difference between the two two-planet models, one with large and one with smaller prior. All the values are given as the average and dispersion calculated for all the simulated data in each scenario.}
         \scriptsize
       \centering
        \begin{tabular}{lccccccccc}
           \hline
            \noalign{\smallskip}
            & &  &  & \multicolumn{6}{c}{two-planet model, narrow prior}\\ 
           Jump  parameter& \cite{damasso19} & Prior &Figure of merit & H19 & H19-23 & H19+U20-23 & H19+All 20-23 & H19-23 RV$^{+}$ & H19-21\\
             \noalign{\smallskip}
             \hline
            \noalign{\smallskip}
            \hline
            \noalign{\smallskip}
%            \begin{center}
            \textbf{GP hyper-parameters} \\
%            \end{center}
            \hline
            \noalign{\smallskip}
            $h$ [m$\,s^{-1}$] & 1.7$_{\rm -0.2}^{\rm +0.3}$ & $\mathcal{U}$(0,4) & $h_{\rm ratio}$ & 1.01$_{\rm -0.06}^{\rm +0.16}$ & 1.00$_{\rm -0.13}^{\rm +0.07}$ & 1.00$_{\rm -0.14}^{\rm +0.09}$ & 0.94$_{-0.12}^{+0.10}$ & 1.0$\pm$0.1 & 0.98$_{-0.10}^{+0.25}$ \\
            \noalign{\smallskip}
            $\lambda$ [days] & 398$^{\rm +122}_{\rm -279}$ & $\mathcal{U}$(0,1000) & $\lambda_{\rm ratio}$ & 1.0$\pm$0.3 & 0.92$^{+0.09}_{-0.15}$ & 0.95$_{\rm -0.19}^{\rm +0.15}$ & 0.93$^{+0.09}_{-0.15}$ & 0.96$_{-0.08}^{+0.20}$ & 1.09$_{-0.18}^{+0.08}$\\
            \noalign{\smallskip}
            $w$  & 0.30$_{\rm -0.05}^{\rm +0.06}$ & $\mathcal{U}$(0,1) & $w_{\rm ratio}$ & 1.1$_{\rm -0.2}^{\rm +0.4}$ & 1.0$_{\rm -0.09}^{\rm +0.1}$ & 0.9$_{\rm -0.1}^{\rm +0.2}$ & 1.0$_{\rm -0.1}^{\rm +0.2}$ & 1.0$\pm$0.1 & 1.0$_{\rm -0.1}^{\rm +0.2}$ \\
            \noalign{\smallskip}
            $P_{\rm rot}$ [days] & 87.8$_{\rm -0.8}^{\rm +0.6}$ & $\mathcal{U}$(80,95) & $P_{\rm rot,\:ratio}$ & 1.001$_{\rm -0.009}^{\rm +0.007}$ & 1.004$_{\rm -0.006}^{\rm +0.004}$ &  0.999$_{\rm -0.004}^{\rm +0.006}$ & 1.000$^{+0.002}_{-0.006}$ & 1.000$_{-0.006}^{+0.005}$ & 0.999$_{-0.002}^{+0.004}$ \\
            \noalign{\smallskip}
            \hline
            \noalign{\smallskip}
            \textbf{Planetary parameters} \\
            \noalign{\smallskip}
            \hline
            \noalign{\smallskip}
            $K_{\rm b}$ [m$\,s^{-1}$] & 1.2$\pm$0.1 & $\mathcal{U}$(0,3) & & & & \\
            \noalign{\smallskip}
            $P_{\rm b}$ [days] & 11.185$\pm$0.001 & $\mathcal{U}$(10.5,12) & & & & \\ 
            \noalign{\smallskip}
            $T_{\rm b,\:conj}$ [BJD-2450000] & 7897.9$_{\rm -0.2}^{\rm +0.3}$ & $\mathcal{U}$(7895,7910) & & & & \\
            \noalign{\smallskip}
            $K_{\rm c}$ [m$\,s^{-1}$] & 1.2$\pm$0.4 & $\mathcal{U}$(0,3)\tnote{a} & $K_{\rm c,\:ratio 50\%}$ & 1.05$_{\rm -0.05}^{\rm +0.08}$ & 1.02$^{+0.19}_{-0.09}$ & 0.96$_{\rm -0.13}^{\rm +0.04}$ & 0.88$^{+0.06}_{-0.10}$ & 1.05$_{-0.09}^{+0.07}$ & 0.98$_{-0.24}^{+0.10}$ \\ 
            \noalign{\smallskip}
            &  & & $K_{\rm c,\:50\%}$/$\sigma^{\rm -}_{\rm K_{\rm c}}$  & 3.24$_{\rm -0.15}^{\rm +0.24}$ & 3.9$^{+0.9}_{-0.5}$ & 3.6$_{\rm -0.3}^{\rm +0.6}$ & 3.4$^{+0.8}_{-0.6}$ & 4.0$_{-0.6}^{+0.5}$ & 3.5$_{-1.1}^{+0.6}$\\
            \noalign{\smallskip}
            $P_{\rm c}$ [days] & 1900$^{\rm +96}_{\rm -82}$ & $\mathcal{U}$(1600,2200)\tnote{b} & $P_{\rm c,\:ratio 50\%}$ & 1.02$_{\rm -0.02}^{\rm +0.01}$ & 1.00$^{+0.01}_{-0.02}$ &  1.01$\pm$0.01 & 1.02$\pm$0.02 & 1.01$_{-0.01}^{+0.02}$ & 1.01$\pm$0.02 \\ 
            \noalign{\smallskip}
            &  & & $P_{\rm c,\:50\%}$/$\sigma^{\rm -}_{\rm P_{\rm c}}$ & 25$_{\rm -1}^{\rm +2}$ & 31$^{+8}_{-4}$ & 30$\pm$4 & 27$^{+6}_{-3}$ & 32$\pm$4 & 28$_{-10}^{+3}$\\
            \noalign{\smallskip}
            $T_{\rm c,\:conj}$ [BJD-2450000] & 5892$^{\rm +101}_{\rm -102}$ & $\mathcal{U}$(5000,7300)\tnote{c} & & & & \\ 
            \noalign{\smallskip}
             \hline
            \noalign{\smallskip}
            \textbf{Bayesian evidence difference} & & & &  &  &  &  &  & \\
            \noalign{\smallskip}
            \hline
            \noalign{\smallskip}
            % $\ln \mathcal{Z}$ & -- & \bf-580.51$\pm$0.06 & & & & \\
            % \noalign{\smallskip}
             $\ln \mathcal{Z}(2p_w)$ - $\ln \mathcal{Z}(1p)$ &  -5.8 &  & & -0.6$\pm$0.5 & 0.03$\pm$1.29 & -0.6$\pm$0.8 & -2.3$\pm$2.7 & 0.4$\pm$1.6 & -0.5$\pm$0.4 \\
             \noalign{\smallskip}
             $\ln \mathcal{Z}(2p_n)$ - $\ln \mathcal{Z}(1p)$ & +1.6 &  & & 1.9$\pm$0.5 & 3.0$\pm$1.5 & 2.2$\pm$1.0 & 2.2$\pm$1.2 & 3.2$\pm$1.4 & 1.9$\pm$1.2\\
             \noalign{\smallskip}
             $\ln \mathcal{Z}(2p_w)$ - $\ln \mathcal{Z}(2p_n)$ &  -7.4 &  & & -2.5$\pm$0.5 & -3.0$\pm$0.4  & -2.8$\pm$0.4 & -4.4$\pm$2.8 & -2.8$\pm$0.8 & -2.5$\pm$0.8  \\
            \noalign{\smallskip}
            \hline
            \hline
\end{tabular} 
\begin{tablenotes}
\item[a] Larger prior: $\mathcal{U}$(0,5)
\item[b] Larger prior: $\mathcal{U}$(20,6500)
\item[c] Larger prior: $\mathcal{U}$(2000,9500)
\end{tablenotes}
\label{tab:sim2planets}
%\end{threeparttable}
\end{table}
\end{landscape}
%\end{scriptsize}     

\begin{table*}
       \caption{Same as Table \ref{tab:sim2planets}, but for \textit{Problem 2}, that means the mock RVs include only the signal of Proxima\,b (circular orbit) and stellar noise calculated in Paper I. They were generated from the parameter best-fit values indicated in column 2. The figures of merit related to planet c are calculated with respect to the injected solution used for \textit{Problem 1}.}
    %     \scriptsize
       \centering
       \tiny 
        \begin{tabular}{llccccc}
           \hline
            \noalign{\smallskip}
           Jump parameter & Model: 1 circular planet & Prior & Figure of merit & H19-23 & H19+U20-23 & H19+All 20-23  \\
           & \cite{damasso19} & (GP+ two planets regression) & (GP+ two planets regression) \\
             \noalign{\smallskip}
             \hline
            \noalign{\smallskip}
            \hline
            \noalign{\smallskip}
%            \begin{center}
            \textbf{GP hyper-parameters} \\
%            \end{center}
            \hline
            \noalign{\smallskip}
            $h$ [m$\,s^{-1}$]  & 1.9$_{\rm -0.2}^{\rm +0.3}$ & $\mathcal{U}$(0,4) & $h_{\rm ratio}$ & 0.9$_{-0.1}^{+0.2}$ & 0.9$\pm$0.1 & 0.9$_{-0.1}^{+0.2}$ \\
            \noalign{\smallskip}
            $\lambda$ [days] & 291$^{\rm +127}_{\rm -243}$ & $\mathcal{U}$(0,1000) & $\lambda_{\rm ratio}$ & 1.1$\pm$0.3 & 1.0$\pm$0.2 & 1.1$_{-0.1}^{+0.3}$ \\
            \noalign{\smallskip}
            $w$ & 0.35$_{\rm -0.06}^{\rm +0.08}$ & $\mathcal{U}$(0,1) & $w_{\rm ratio}$ & 1.0$_{-0.1}^{+0.2}$ & 0.8$_{-0.1}^{+0.3}$ & 0.9$_{-0.1}^{+0.2}$\\ 
            \noalign{\smallskip}
            $\theta$ [days] & 87.8$_{\rm -1.9}^{\rm +1.2}$ & $\mathcal{U}$(80,95) & $P_{\rm rot,\: ratio}$ & 0.99$_{-0.01}^{+0.02}$ & 0.99$\pm$0.02 & 1.000$_{-0.008}^{+0.009}$\\
            \noalign{\smallskip}
            \hline
            \noalign{\smallskip}
            \textbf{Planetary parameters} \\
            \noalign{\smallskip}
            \hline
            \noalign{\smallskip}
            $K_{\rm b}$ [m$\,s^{-1}$] & 1.3$\pm$0.1 & $\mathcal{U}$(0,3) & & &  \\
            \noalign{\smallskip}
            $P_{\rm b}$ [days] & 11.1847$\pm$0.0009 & $\mathcal{U}$(10.5,12) & & &  \\ 
            \noalign{\smallskip}
            $T_{\rm b,\:conj}$ [BJD-2450000] & 7897.94$\pm$0.24 & $\mathcal{U}$(7895,7910) & & &  \\
            \noalign{\smallskip}
            $K_{\rm c}$ [m$\,s^{-1}$] & & $\mathcal{U}$(0,3) & $K_{\rm c,\:ratio 50\%}$ & 0.7$^{+0.2}_{-0.1}$ & 0.7$^{+0.2}_{-0.1}$ & 0.6$^{+0.3}_{-0.1}$ \\ \noalign{\smallskip}
            &  & & $K_{\rm c,\:50\%}$/$\sigma^{\rm -}_{\rm K_{\rm c}}$ & 1.9$^{+1.1}_{-0.2}$ & 2.4$^{+1.0}_{-0.5}$ & 2.1$^{+1.1}_{-0.6}$ \\
            \noalign{\smallskip}
            $P_{\rm c}$ [days] & & $\mathcal{U}$(1600,2200) & $P_{\rm c,\:ratio 50\%}$ & 0.99$\pm$0.06 & 1.02$\pm$0.05 & 1.03$\pm$0.06 \\
            \noalign{\smallskip}
            &  & & $P_{\rm c,\:50\%}$/$\sigma^{\rm -}_{\rm P_{\rm c}}$ & 20$_{-9}^{+4}$ & 19$_{-4}^{+9}$ & 21$_{-10}^{+7}$\\
            \noalign{\smallskip}
            $T_{\rm c,\:conj}$ & & $\mathcal{U}$(5000,7300) & & & \\ 
            \hline
            \noalign{\smallskip}
            \textbf{Bayesian evidence difference} \\
            \noalign{\smallskip}
            \hline
            \noalign{\smallskip}
       %      $\ln \mathcal{Z}$ & -- & \bf-582.1$\pm$0.03 & & & \\
        %     \noalign{\smallskip}
            % $\ln \mathcal{Z}(2p_w) - \ln \mathcal{Z}(1p)$ & -- & \bf -5. 8 &  \bf & &  \\
            % \noalign{\smallskip}
             $\ln \mathcal{Z}_{2 \:planets} - \ln \mathcal{Z}_{1 \:planet}$ & +1.6 &  & & -0.08$\pm$1.16 & 0.3$\pm$1.2 & 0.0$\pm$1.1 \\
            % \noalign{\smallskip}
           %  $\ln \mathcal{Z}(2p_w) - \ln \mathcal{Z}(2p_n)$&-- & \bf -7.4 & & &  \\
            \noalign{\smallskip}
            \hline
            \hline
\end{tabular}\label{tab:sim1planet}
\end{table*}

\section{Conclusions}
\label{sec:conclusions}
We investigated the expectations of a long-term radial velocity follow-up concerning the existence of Proxima\,c, and quantified how much the planet detection significance could improve after collecting more data. We analysed time series spanning $\sim$23-year, containing simulated RVs for the period 2019-2023 added to the $\sim$17-year-long archival data. We consider measurements from the high-resolution spectrogpraphs HARPS and UVES, and simulated different observing strategies we deem affordable. %, so excluding the case of a very dense sampling over 5 years, which does not seem realistic. 
The results were derived using the same analysis tools and set-up adopted in \cite{damasso19}.
Our main conclusion is that 5 years of observations, approximately equivalent to one orbit of Proxima\,c, should not bring the detection above a 4$\sigma$ significance, nor will strongly improve the Bayesian evidence of the two-planet model with respect to the one-planet model. Meanwhile, a confirmation, or disprove, can arrive from astrometry with the Gaia satellite. At the same time, in the hypothesis that only Proxima\,b exists, we show that a signal with about the predicted orbital period of Proxima\,c still shows up, but its average statistical significance substantially lowers. We then forecast it will be challenging also to completely disprove the existence of Proxima\,c with only 5 more years of RVs.
In the future it will be interesting to extend this approach to mock data taken with ESPRESSO, as soon as the first real data will become available. The expected high performances of ESPRESSO, in terms of accuracy and precision, could improve the detection and characterization of the signal of Proxima\,c without the need of a dense sampling. 

\section*{Acknowledgements}
We thank the anonymous referee for the helpful comments.
      M.D. acknowledges financial support from Progetto Premiale 2015 FRONTIERA (OB.FU. 1.05.06.11) funding scheme of the Italian Ministry of Education, University, and Research. 
      F.D.S. gratefully acknowledges the support and the hospitality of
INAF Astrophysical Observatory of Torino, and the Project HPC-EUROPA3 (INFRAIA-2016-1-730897),
with the support of the EC Research Innovation Action under the H2020 Programme.
We acknowledge the computing centres of INAF - Osservatorio Astronomico di Trieste / Osservatorio Astrofisico di Catania, under the coordination of the CHIPP project, for the availability of computing resources and support. We thank A. Celi, G. Moschin, U. Tognazzi, P. Noiret, and D. Del Prete for a very significant long-term friendship.

%%%%%%%%%%%%%%%%%%%%%%%%%%%%%%%%%%%%%%%%%%%%%%%%%%



\bibliographystyle{mnras}
\bibliography{biblio.bib} % your references Yourfile.bib

\begin{thebibliography}{}
\makeatletter
\relax
\def\mn@urlcharsother{\let\do\@makeother \do\$\do\&\do\#\do\^\do\_\do\%\do\~}
\def\mn@doi{\begingroup\mn@urlcharsother \@ifnextchar [ {\mn@doi@}
  {\mn@doi@[]}}
\def\mn@doi@[#1]#2{\def\@tempa{#1}\ifx\@tempa\@empty \href
  {http://dx.doi.org/#2} {doi:#2}\else \href {http://dx.doi.org/#2} {#1}\fi
  \endgroup}
\def\mn@eprint#1#2{\mn@eprint@#1:#2::\@nil}
\def\mn@eprint@arXiv#1{\href {http://arxiv.org/abs/#1} {{\tt arXiv:#1}}}
\def\mn@eprint@dblp#1{\href {http://dblp.uni-trier.de/rec/bibtex/#1.xml}
  {dblp:#1}}
\def\mn@eprint@#1:#2:#3:#4\@nil{\def\@tempa {#1}\def\@tempb {#2}\def\@tempc
  {#3}\ifx \@tempc \@empty \let \@tempc \@tempb \let \@tempb \@tempa \fi \ifx
  \@tempb \@empty \def\@tempb {arXiv}\fi \@ifundefined
  {mn@eprint@\@tempb}{\@tempb:\@tempc}{\expandafter \expandafter \csname
  mn@eprint@\@tempb\endcsname \expandafter{\@tempc}}}

\bibitem[\protect\citeauthoryear{{Ambikasaran}, {Foreman-Mackey}, {Greengard},
  {Hogg}  \& {O'Neil}}{{Ambikasaran} et~al.}{2015}]{george15}
{Ambikasaran} S.,  {Foreman-Mackey} D.,  {Greengard} L.,  {Hogg} D.~W.,
  {O'Neil} M.,  2015, \mn@doi [IEEE Transactions on Pattern Analysis and
  Machine Intelligence] {10.1109/TPAMI.2015.2448083}, \href
  {https://ui.adsabs.harvard.edu/abs/2015ITPAM..38..252A} {38, 252}

\bibitem[\protect\citeauthoryear{{Anglada-Escud{\'e}}
  et~al.,}{{Anglada-Escud{\'e}} et~al.}{2016}]{anglada16}
{Anglada-Escud{\'e}} G.,  et~al., 2016, \mn@doi [\nat] {10.1038/nature19106},
  \href {https://ui.adsabs.harvard.edu/abs/2016Natur.536..437A} {536, 437}

\bibitem[\protect\citeauthoryear{{Buchner} et~al.,}{{Buchner}
  et~al.}{2014}]{buchner14}
{Buchner} J.,  et~al., 2014, \mn@doi [\aap] {10.1051/0004-6361/201322971},
  \href {http://adsabs.harvard.edu/abs/2014A%26A...564A.125B} {564, A125}

\bibitem[\protect\citeauthoryear{{Cloutier}, {Doyon}, {Menou}, {Delfosse},
  {Dumusque}  \& {Artigau}}{{Cloutier} et~al.}{2017}]{cloutier17}
{Cloutier} R.,  {Doyon} R.,  {Menou} K.,  {Delfosse} X.,  {Dumusque} X.,
  {Artigau} {\'E}.,  2017, \mn@doi [\aj] {10.3847/1538-3881/153/1/9}, \href
  {https://ui.adsabs.harvard.edu/abs/2017AJ....153....9C} {153, 9}

\bibitem[\protect\citeauthoryear{{Damasso} et~al.,}{{Damasso}
  et~al.}{2018}]{damasso18}
{Damasso} M.,  et~al., 2018, \mn@doi [\aap] {10.1051/0004-6361/201732459},
  \href {https://ui.adsabs.harvard.edu/abs/2018A&A...615A..69D} {615, A69}

\bibitem[\protect\citeauthoryear{{Damasso}, {Pinamonti}, {Scandariato}  \&
  {Sozzetti}}{{Damasso} et~al.}{2019}]{2019DPSS}
{Damasso} M.,  {Pinamonti} M.,  {Scandariato} G.,   {Sozzetti} A.,  2019,
  \mn@doi [\mnras] {10.1093/mnras/stz2216}, \href
  {https://ui.adsabs.harvard.edu/abs/2019MNRAS.489.2555D} {489, 2555}

\bibitem[\protect\citeauthoryear{{Damasso} et~al.,}{{Damasso}
  et~al.}{2020}]{damasso19}
{Damasso} M.,  et~al., 2020, \mn@doi [Science Advances]
  {10.1126/sciadv.aax7467}, \href
  {https://ui.adsabs.harvard.edu/abs/2020SciA....6.7467D} {6, eaax7467}

\bibitem[\protect\citeauthoryear{Feroz, Balan  \& Hobson}{Feroz
  et~al.}{2011}]{feroz11}
Feroz F.,  Balan S.~T.,   Hobson M.~P.,  2011, \mn@doi [Monthly Notices of the
  Royal Astronomical Society] {10.1111/j.1365-2966.2011.18962.x}, 415, 3462

\bibitem[\protect\citeauthoryear{{Feroz}, {Hobson}, {Cameron}  \&
  {Pettitt}}{{Feroz} et~al.}{2013}]{feroz13}
{Feroz} F.,  {Hobson} M.~P.,  {Cameron} E.,   {Pettitt} A.~N.,  2013, arXiv
  e-prints, \href {https://ui.adsabs.harvard.edu/abs/2013arXiv1306.2144F} {p.
  arXiv:1306.2144}

\bibitem[\protect\citeauthoryear{{Klein} \& {Donati}}{{Klein} \&
  {Donati}}{2019}]{klein19}
{Klein} B.,  {Donati} J.~F.,  2019, \mn@doi [\mnras] {10.1093/mnras/stz1953},
  \href {https://ui.adsabs.harvard.edu/abs/2019MNRAS.488.5114K} {488, 5114}

\bibitem[\protect\citeauthoryear{{Klein} \& {Donati}}{{Klein} \&
  {Donati}}{2020}]{KD2020MNRAS}
{Klein} B.,  {Donati} J.~F.,  2020, \mn@doi [\mnras] {10.1093/mnrasl/slaa009},
  \href {https://ui.adsabs.harvard.edu/abs/2020MNRAS.493L..92K} {493, L92}

\bibitem[\protect\citeauthoryear{{Nava}, {L{\'o}pez-Morales}, {Haywood}  \&
  {Giles}}{{Nava} et~al.}{2020}]{N2020AJ}
{Nava} C.,  {L{\'o}pez-Morales} M.,  {Haywood} R.~D.,   {Giles} H. A.~C.,
  2020, \mn@doi [\aj] {10.3847/1538-3881/ab53ec}, \href
  {https://ui.adsabs.harvard.edu/abs/2020AJ....159...23N} {159, 23}

\bibitem[\protect\citeauthoryear{{Zechmeister} \& {K{\"u}rster}}{{Zechmeister}
  \& {K{\"u}rster}}{2009}]{zech09}
{Zechmeister} M.,  {K{\"u}rster} M.,  2009, \mn@doi [\aap]
  {10.1051/0004-6361:200811296}, \href
  {http://adsabs.harvard.edu/abs/2009A%26A...496..577Z} {496, 577}

\makeatother
\end{thebibliography}
%
%-------------------------------------------------------------------

% Don't change these lines
\bsp	% typesetting comment
\label{lastpage}
\end{document}